\documentclass[%
 reprint,
superscriptaddress,
nofootinbib,
 amsmath,amssymb,
 aps,
]{revtex4-1}

\usepackage{ marvosym }
\usepackage{graphicx}
\usepackage{dcolumn}
\usepackage{bm}
\usepackage{url}
\usepackage{xcolor}

\begin{document}

\preprint{}

\title{Reply to a comment on `Understanding the $\gamma$-ray emission from the globular cluster 47 Tuc: evidence for dark matter?' }

\author{Anthony M. Brown}\email{anthony.brown@durham.ac.uk}
\affiliation{Centre for Advanced Instrumentation, Department of Physics, University of Durham, South Road, Durham, DH1 3LE, UK}

\author{Thomas Lacroix}
\affiliation{Laboratoire Univers \& Particules de Montpellier (LUPM), Universit\'{e} de Montpellier, CNRS, Universit\'{e} de Montpellier, Montpellier, France}
\author{Sheridan Lloyd}
\affiliation{Centre for Advanced Instrumentation, Department of Physics, University of Durham, South Road, Durham, DH1 3LE, UK}
\author{C\'eline B\oe hm}
\affiliation{School of Physics, University of Sydney, Camperdown, NSW 2006, Australia}
\affiliation{Institute for Particle Physics Phenomenology, Durham University, South Road, Durham, DH1 3LE, United Kingdom}
\affiliation{LAPTH, U. de Savoie, CNRS,  BP 110, 74941 Annecy-Le-Vieux, France}
\affiliation{Perimeter Institute, 31 Caroline St N., Waterloo Ontario, Canada N2L 2Y5}
\author{Paula Chadwick}
\affiliation{Centre for Advanced Instrumentation, Department of Physics, University of Durham, South Road, Durham, DH1 3LE, UK}

\date{\today}

\begin{abstract}
Analysing 9 years of \textit{Fermi}-LAT observations, we recently studied the spectral properties of the prominent globular cluster 47 Tuc \cite{brown47tuc}. In particular, we investigated several models to explain the observed gamma-ray emission, ranging from millisecond pulsars (MSP) to Dark Matter (DM) \cite{brown47tuc}, with the motivation for the latter model driven by recent evidence that 47 Tuc harbours an intermediate-mass black hole (IMBH; \cite{imbh}). This investigation found evidence that the observed gamma-ray emission from 47 Tuc is due to two source populations of MSPs and DM. In \cite{bartels}, the authors comment that this evidence is an artifact of the MSP spectra used in \cite{brown47tuc}. Here we reply to this comment and argue that the authors of \cite{bartels} (i) do not give due consideration to a very important implication of their result and (ii) there is tension between our MSP fit and their MSP fit when taking uncertainties into consideration. As such, we still conclude there is evidence for a DM component which motivates a deeper radio study of the prominent globular cluster 47 Tuc. 
\end{abstract}

\maketitle

\section{Introduction}
47 Tuc was the first globular cluster found to be gamma-ray bright \cite{47tucfermiscience}, with the gamma-ray emission being attributed to an unresolved population of millisecond pulars (MSPs). 47 Tuc is also one of the few globular clusters that might harbour an IMBH \cite{imbh}. The presence of such an object within 47 Tuc leads us to consider the possibility that some of the observed gamma-ray emission from 47 Tuc can be attributed to the by-products of annihilating dark matter (DM), as the IMBH could enhance the DM density in its close environ \cite{gondolo,horiuchi}. 

With 9 years of \textit{Fermi}-LAT observations, we previously investigated the spectral properties of 47 Tuc with unprecedent accuracy and sensitivity \cite{brown47tuc}. The increased exposure of our study, compared to earlier studies, discovered significant emission below 200 MeV. To investigate the origin of the observed gamma-ray emission, we conducted detailed spectral modelling. 47 Tuc has 25 resolved MSPs \cite{freire}. To account for this source of gamma-rays, we assumed the previously published MSP spectrum of Xing \& Wang \cite{xingwang}. This spectrum was derived by simultaneously fitting the normalised spectrum of 39 out of the 40 MSPs within \textit{Fermi}-LAT's 2nd pulsar catalogue (2PC; \cite{2pc}). To account for the large variance in the spectra of the MSP population of the 2PC, Xing \& Wang considered a systematic uncertainty parameter that was added in quadrature to each spectral bin, for each MSP. The best-fit spectral shape was a power-law with an exponential cut-off, a spectral index of $\Gamma=1.54^{+0.10}_{-0.11}$ and a cut-off energy of $E_c = 3.70^{+0.95}_{-0.70}$ GeV \cite{xingwang}, with the uncertainties of these parameters representing $3\sigma$ uncertainties. To account for any possible gamma-ray emission from DM, we considered a spike in the DM density in the immediate vicinity of an IMBH within 47 Tuc, with the radius of the spike density being set by the mass of the IMBH \cite{brown47tuc} \footnote{This description of DM within the vicinity of a black hole has been successfully applied to other astrophysical systems (e.g. \cite{brown}).}. For the spectral fit, a maximum likelihood analysis was considered, with the DM mass and annihilation cross-section being treated as free parameters. Considering these two population descriptions, we found that a two-source `MSP$+$DM' description of 47 Tuc's spectrum was preferred over a `MSP-only' description with a test-statistic difference of TS$=$40. 

Ref \cite{bartels} disputes the conclusions of our paper. In particular, the authors question the MSP spectral description that we assumed, arguing that it does not take into consideration the variance in the spectral shapes of the MSP population within the 2PC. Instead, \cite{bartels} uses their own bespoke MSP spectral model using a synthetic mock MSP catalogue derived using the luminosity function of disk MSPs. 

In this paper, we discuss two key areas \cite{bartels}: the spectral model they assume and the implications of their conclusions with respect to pulsed gamma-ray emission. Our arguments on these points bring the conclusions of \cite{bartels} into question. We do offer an alternative MSP-based argument that counters the weaknesses of \cite{bartels}, although there is as yet no observational evidence to support this alternative. As such, we feel that there is still sufficient evidence to warrant the consideration of DM within 47 Tuc. 

\section{Discussion}

\subsection{Uncertainties in the assumed MSP description.}
The authors of ref \cite{bartels} proposed an alternative bespoke MSP spectral description on the assumption that the approach of Xing \& Wang did not take into consideration the variance in the spectral shapes of the 2PC's MSP population. As discussed in the introduction above, this assumption is not correct, with the reported uncertainties on the spectral index and cut-off energy of their model being at a conservative $3\sigma$ level to account for this spectral variance. This $3\sigma$ uncertainty was not taken into consideration when conducting the original model fitting in \cite{brown47tuc}. We have performed additional maximum likelihood fits for the `MSP-only' model, with the spectral index of our assumed MSP spectral model fixed to the extreme values allowed by the $3\sigma$ uncertainties; ie $\Gamma_{low}=1.43$ and $\Gamma_{high}=1.64$. Comparing the log-likelihood of these fits to that of the `MSP$+$DM' two-source population fits of \cite{brown47tuc}, we find TS values of TS$_{low}=21$ and TS$_{high}=92$ respectively. For one degree of freedom, this equates to significances of $4.6\sigma$ to $9.6 \sigma$ respectively.

The range of TS values of these fits has two important implications. Firstly, the significance of the MSP fit is sensitive to the spectral index of the MSP population. Coupling this sensitivity with a large uncertainty in the index, as is the case for the $3\sigma$ uncertainty of the Xing \& Wang MSP model used here, may result in spurious signals being deemed to be significant. Mitigating against this requires a more accurate MSP model.

Secondly, the range of TS values aside, we note that for the extreme hard spectral index case, $\Gamma_{low}=1.43$, the TS value drops to a level that is below the $5\sigma$ discovery threshold, and as such, we are unable to state that there is a significant preference for a two-source model. Nonetheless, even at this extreme index value, the TS value of the two-source model (TS$_{low}=21$) is still large enough to warrant the suggestion that there is a preference for the two-source model when compared to an MSP only model. This is clearly at tension with \cite{bartels}'s statement that once the variance in spectral shapes is taken into consideration, there is no difference in likelihood between a one and two-source population description of 47 Tuc's gamma-ray emission. 

To investigate the reason for this discrepancy requires us to compare, in detail, how both MSP models are derived. The MSP model assumed by \cite{brown47tuc}, including uncertainties, has previously been published in a refereed journal, with the derivation of this model being open to investigation by the wider scientific community. The MSP model proposed by \cite{bartels} is a bespoke synthetic model, derived specifically for 47 Tuc, from a previously published luminosity function. This derivation is neither published or outlined sufficiently in \cite{bartels}, and as such, we are unable to investigate the discrepancy.

\subsection{Pulsed gamma-ray emission}
The most obvious `test-able' prediction of \cite{bartels}'s derived MSP model is the presence of pulsed gamma-ray emission. Ref \cite{bartels} claims that typically half of 47 Tuc's flux can be attributed to 5 MSPs\footnote{In private communications the authors of \cite{bartels} acknowledged that there were instances where their mock spectra had significant contribution from just one MSP.}. A consequence of the gamma-ray flux being dominated by a small number of bright MSPs would be the presence of gamma-ray pulsations in 47 Tuc's gamma-ray flux. Previous studies have found no evidence of such pulsation \cite{47tucfermiscience}. 

While \cite{bartels} claims that typically half of 47 Tuc's flux can be attributed to 5 MSPs, when addressing the possibility of pulsed gamma-ray emission, \cite{bartels} refers to the atypical instance that 47 Tuc's gamma-ray flux is dominated by 10 MSPs, and simply states that detecting pulsations against a large background will become difficult. Contrary to this statement by \cite{bartels}, the Einstein\MVAt home gamma-ray pulsar survey project, which is a blind survey, has discovered pulsed gamma-ray emission from numerous MSPs \cite{einstein,einsteinweb}. Importantly a large percentage of these MSPs is located in the area with the most luminous diffuse emission on the sky, the Galactic Bulge, where the gamma-ray luminosity is $(3.9 \pm 0.5) \times 10^{36}$ ergs s$^{-1}$ (eg. \cite{oscar}). All bar one of the newly discovered pulsed MSP located in the Galactic bulge have spin-down luminosities in the range of $10^{34}$ to $10^{35}$ ergs s$^{-1}$. Assuming a conservative spin-down luminosity to gamma-ray luminosity conversion factor of 10\% \cite{47tucfermiscience}, this pulsed gamma-ray emission is 0.04\% to 0.74\% of the diffuse emission in which they are embedded\footnote{Assuming an unrealistic 100\% spin-down to gamma-ray luminosity conversion, the pulsed emission is still 0.4 to 7.4\% of the diffuse emission, and as such, by extension that \cite{bartels} statement that 10 pulsars are responsible for the majority of the gamma-ray flux is unrealistic.}. This observational evidence is contrary to \cite{bartels} claim that the detection of pulsed emission from an MSP with a luminosity 5\% of 47 Tuc's would not be possible.

Outside of the Galactic bulge, the faintest pulsed gamma-ray emitting MSP found by Einstein\MVAt home has a gamma-ray luminosity of $3.7 \times 10^{32}$ ergs s$^{-1}$ (again assuming at 10\% conversion efficiency), a factor of 20 times fainter than 47 Tuc's gamma-ray luminosity \cite{brown47tuc}. Again, this observational evidence is contrary to the claim by the authors of \cite{bartels} that the detection of pulsed emission from an MSP with a luminosity 5\% of 47 Tuc's would not be possible.   

\section{Conclusions}

We address two key aspects of \cite{bartels}'s comment on our recent work on 47 Tuc \cite{brown47tuc}. In particular, we discuss two key areas of \cite{bartels}'s work: the spectral model they assume and the implications of their conclusions with respect to pulsed gamma-ray emission. To account for \cite{bartels}'s concerns that our preference for a two-source model was based on an artifact of the variance in spectral models within the 2PC's MSP population, we performed additional likelihood fits, with extreme MSP spectral parameters which accounted for this variance. Even with an extremely hard index, the likelihood fit has a TS$=21$, indicating that these additional fits still find strong evidence suggesting a two-source model is preferred over an MSP only model. With regards to \cite{bartels} dismissing the possibility of pulsed gamma-ray emission being too difficult to find, we provide published observational evidence to the contrary, citing several instances where (i) pulsed gamma-ray emission has been observed from MSPs embedded in a strong diffuse flux and (ii) pulsed emission has been observed from faint MSPs, 5\% the luminosity of 47 Tuc. 

As such, we feel that there is still evidence that the gamma-ray emission from 47 Tuc is potentially due to two source populations: annihilation DM and an ensemble of MSPs. We do however note that there is an alternative explanation that neither \cite{brown47tuc} or \cite{bartels} has considered: that the gamma-ray emission from 47 Tuc is due to a sizeable population of MSPs at the faint end of the MSP luminosity function (see \cite{harding}). Such a population would have a harder spectral index, which is more compatible with the observed gamma-ray spectrum of 47 Tuc, and would not exhibit pulsed gamma-ray emission. To test this alternative explanation requires deep radio observations of 47 Tuc. We strongly encourage that these observations be performed.

\section{Acknowledgements}
AMB would like to acknowledge the fruitful conversation with Alice Harding during the 2018 \textit{Fermi} symposium in Baltimore.

\end{document}